# Reliable Web Services Approaches of Mobile Cloud Computing: A Comparative study

Amr S.Abdelfattah
Computer Science Department,
Faculty of Computers & Information Sciences, Ain-Shams University
amr.elsayed@cis.asu.edu.eg

Tamer Abdelkader
Information Systems Department
Faculty of Computers & Information Sciences, Ain-Shams University
tammabde@cis.asu.edu.eg

EI-Sayed M. EI-Horbaty
Computer Science Department
Faculty of Computers & Information Sciences, Ain Shams University
Shorbaty@cis.asu.edu.eg

*Abstract*— The Mobile intermittent wireless connectivity limits the evolution of the mobile landscape, such that this progress coupled with the ubiquitous nature of the Internet and the web services. The web service Reliability achieves the low overhead communication and retrieves the appropriate state response from the web service consumption. This paper discusses and analyzes the most recent approaches that achieve the reliability of web services consumed by mobile devices. Achieving the reliable web service consumption is tended to two approaches: Middleware approach and Mobile Agent (MA) approach. Both of them focus on ensuring the request execution under the communication limitations and services temporary unavailability.

*Keywords—Web service; Reliable request; Mobile Cloud Computing.*

## I. Introduction

The evolution of the mobile landscape coupled with the ubiquitous nature of the Internet and the recent explosion of the cloud computing technology is facilitating the deployment of web services. The web services are the perfect way to provide a standard platform and operating system independent mechanism for enterprise and personalized mobile applications communication over the Web.

Smart phones are gradually becoming the effective client platform to consume the services and the pool of data and information. This is because cloud computing improves scalability and consistency of services and data, and facilitates the deployment of enterprise and personalized mobile applications [1].

The Mobiles "thin clients" are uncomfortable and expensive in terms of time and effort. Because of their limited processing power and intermittent wireless connectivity, there is uncertainty of whether the web service request was successfully received, was lost in the internet before reaching the server, or was partially processed. In the case the application retries the operation and resends the request, it may be duplicated or cause an error, such as two orders entered or two credit card charges.

Web services encapsulate Cloud Computing because Cloud Computing uses web services for connections. Most of these cloud oriented services and data are deployed as web services that are network-oriented applications [2]. The synchronization between a mobile device and a web service is achieved through initiating a conversation in a request response pattern.

Mobile Cloud Computing (MCC) is the combination of cloud computing, mobile computing and wireless networking to bring rich computational resources to mobile users, network operators, as well as cloud computing providers [3].

The reliable web services through mobile cloud computing achieved using approaches that focus on ensuring the request execution under the intermittent connectivity, services unavailability conditions, moreover ensuring the appropriate response according to the request state.

The rest of this paper is organized as follows: In Section II, web services types are discussed. In Section III, the challenges that prevent achieving the reliable request are discussed. In Section IV, CAP Theorem will be explained. In Section V, the main approaches for achieving the reliability are compared. In Section VI, we analyze the results and the usage of the reliable approaches. The conclusion and future work are given in Section XI.

## II. Web services types and limitations

Web services act as self-contained components, which are published, located and invoked over the web. The key concept behind web services is to provide a standard platform and operating system independent mechanism for application communication over the web.

The web service Architecture (WSA) [4] proposed by W3 organization relies on a number of Web standards, such as Simple Object Access Protocol (SOAP), Web Service Description Language (WSDL), XML, and Universal Description Discovery and Integration (UDDI) that allows services to be searched, described and integrated by any application [5]. There are two types of web services: Simple Object Access Protocol (SOAP) and Representational State Transfer (REST).

The two major SOAP-based web services limitations are categorized mainly to communication and computation overhead [5]:

1. Communication overhead

- The verbose nature of the XML-based messaging system was not intended for communication efficiency.
- Communication performance issues due to lack of support for transaction in communication with a web service.
- Communication may suffer poor performance over busy or unstable networks in comparison with other traditional approaches of distributed computing such as CORBA or DCOM.
- It makes this kind of data exchange protocols stateless as the web service provider and web service consumer don't have knowledge of each other's state.

2. Computation overhead

- Encoding/Decoding XML requests/responses definitely degrades the overall application performance.
- Performance issues due to XML processing and parsing overhead.

The REST architecture is fundamentally client-server architecture, and is designed to use a stateless communication protocol, typically HTTP. In the REST architecture, clients and servers exchange representations of resources using a standardized interface and protocol. These principles encourage REST applications to be simple, lightweight, and have high performance. Therefore, regarding the scope of reliability RESTFUL services overcome SOAP services limitations and achieve better results especially in mobile communications [6, 7, 8]. Using REST as a service architecture is preferred for mobile devices because REST services use HTTP request and response, which means that a mobile device connected with the internet can access the service without additional overhead, unlike SOAP web services [9].

In [10], the authors observed that using REST enabled their system to scale with multiple users and devices, but also complained about request-response mechanism, which sometimes limits the system at the network level.

In [11], the authors argue that many firewalls permit only the GET and POST methods. There is also a size limit on the URI for the GET method encoding; and HTTPS is not cacheable [1].

SOAP web services development, their clients require extra effort mostly due to the lack of native support. According to [12], combining RESTful design with other technologies such as caching provides good system scalability.

### III. Reliability and Challenges

The challenges in mobile web services consumption are worth investigating from two perspectives: Mobile limitations, and connectivity limitations from Mobile side and Cloud service provider side. In addition the network bandwidth limitation and unreliable wireless communication are decreasing the overall support for web services consumption on mobile devices [5]. The challenges are listed in the following points:

- Connection loss: Form the perspective of Smart phones, clients have intermittent connection because of their mobility. They can be momentarily removed from the connected network and later join the available network [9]. From the perspective of Cloud/ Server, it may lose the connection and their web services become unreachable from clients.

- latency/ Bandwidth: Mobile Cellular networks have a very limited bandwidth and are often billed based on the amount of data transferred.

- Reliability: The public Internet is unreliable if a client calls a web service and doesn't get a response within the timeout period, the client doesn't know whether the request was received, partially processed, or lost to perform the appropriate action.

- Longer Transaction time: Web service consumption will take longer to execute than an equivalent direct query against a local database. The consumption will be slower because of the HTTP overhead, the XML overhead, and the network overhead to a remote server. Therefore the differences in performance between Web services and traditional database queries need to be factored into the application architecture to prevent unexpectedly poor performance due to the latency of the web services consumption failures.

However, there is even a bigger challenge that has been overlooked recently by researchers within the distributed mobile network which is the "CAP Theorem" [13], discussed in Section IV. In a highly distributed system where web services are scattered across multiple platforms, three system guarantees are required: Consistency of the data, Availability of the system/data, and Partition tolerance to fault. However, the CAP theorem states that at most only two of the three can be guaranteed simultaneously. In distributed mobile systems where the mobile node is employed as the client platform of the web services, partition tolerance is a given because of the intermittent connectivity losses. This means we are forced to choose between Availability and Consistency [1].

### IV. The CAP Theorem

Web services running in scalable systems are expected to provide the following support: Consistency of the data/service, Availability of the system/data, and Partition tolerance. However, In [14], the author makes a conjecture which states that no distributed system can guarantee all three requirements at the same time. It was noted that two of the three requirements can be guaranteed simultaneously if one requirement can be traded-off. This idea is conceptualized below in Figure 1. In [15], the authors use a formal module based on read/write operations on disjointed nodes to prove the conjecture into a theorem.

**Consistency** is the requirement which guarantees that states stored in a distributed system such as enterprise mobile based systems are seen by clients the same at every node [16]. In addition, the authors in [15] describe consistency as "atomic", which means that every system transaction is one and must be completed fully or not started at all. To ensure a consistent system, some developers and researchers have adopted the ACID (Atomicity, Consistency, Isolation, and Durability) technique [17], especially within the context of databases.

- Atomicity: Considering an operation in a distributed system, if a part fails then the entire transaction fails and the system must be rolled back to its original state. This means that each transaction is single and cannot be broken into parts.

- Consistency: This property ensures that anytime a request is issued to the database, the requester receives a valid and desirable data, that means the database state will be consistent when the transaction begins and ends.

- Isolation: A distributed system should be built in a way that transactions being processed should be hidden from other processes until it comes to completion.

- Durability: After a successful transaction, the operation cannot be rolled back even if there are system failures.

Transactions even though atomic require a lot of sequence of operations. Distributed systems normally rely on the two-phase commit (2PC) protocol, the three-phase (3PC) commit protocol, and timestamps to provide the ACID capabilities. In [17], the author explains the two-phase commit (2PC) as follows: If all databases agree on an operation (i.e., precommit) to the coordinator, then the operation can continue with the coordinator in the second phase asking the operations to commit, but if one reports failure, the entire system must be rolled back. Since most systems are horizontally scaled, it means partition tolerance is ensured which means systems cannot be available according to the CAP theorem. In addition, because of the use of database keys, distributed transactions are highly coupled [17]. It is also challenging to build long-lived transactions in ACID since there is locking and serialization. Moreover, the challenge with ensuring consistency is the high cost of bandwidth consumption due to the communication overhead since all the participating nodes have to commit to an atomic transaction [18]. As we discussed in the section III, because mobile devices are not guaranteed reliable connectivity, it is impractical and almost impossible to ensure locking especially when a mobile participant is not available to commit at a beginning of a transaction.

**Availability** ensures that when some nodes in a distributed system become inaccessible as a result of failures, the other nodes should continue to operate and support every read and write operation [16], [18]. It is important that intended responses are received for each request [17]. Availability can be achieved if consistency is compromised when the BASE (Basically Available Soft-state Eventual-consistency) [17] technique is adopted. BASE is applied to minimize the level of coupling in ACID systems. Decoupling transactions is achieved by introducing persistent message queues. This process ignores the two-phase commit and ensures latency.

**Partition-tolerance** is achieved when a distributed system is built to allow arbitrarily loss of sent messages from one node to another [15]. The current demand from web service consumers makes it impractical to keep all data at one source. This is because when the source fails, it means the entire system becomes unavailable. Partition-tolerance therefore allows for system states to be kept in different locations. This property is a given in a mobile distributed environment since mobility and offline access to data can only be achieved through partitioning.

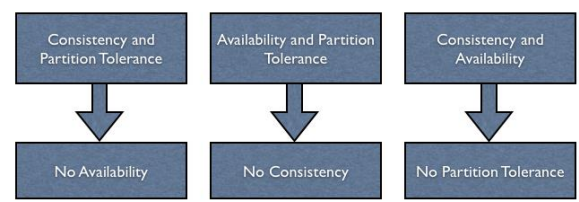

Figure 1: The CAP theorem possibilities

The best pair of guarantees normally preferred by users is Availability and Partition-tolerance while trading-off consistency because it is better to give some data rather than access denied in most workflows [16].

Though the CAP Theorem has been described in many studies in relation to databases, the phenomenon is present in web services and follows the same analysis, Therefore using the theorem concepts will be useful in supporting the reliability in the desired distributed system, which consists of web services and fully available mobile devices nodes consuming these web services.

## V. Reliable Web service Approaches

We can summarize the related work in two approaches:

1. Middleware component that takes the heavy load of the communication with the Web service, as shown in Figure 2.

2. Mobile agent (MA) that represents the mobile user in wireless networks to consume a web service based on user input. Then the MA sends the result if the user is online, otherwise it sends the result after his/her network session is re-establishmed, as shown in Figure 3.

### A. Middleware Approach

The middleware component acts as a gateway that communicates lightly with the client while ensuring the responsibility of retrieving the response from the web service.

The proposed communication architecture introduces a gateway between the mobile client and the web service that takes all the burden of the heavy load communication with the service. The mobile client will instead have to sustain a lightweight and simple client-server communication over a fast binary protocol [3]. The future middleware solution for mobile clients mostly focuses on application and content adaptation.

The general middleware four fundamental requirements are Heterogeneity, Network communication, reliability, and coordination. Scalability can be achieved with distributed middleware. Context can help middleware adapt to the heterogeneous environment. However, the goal of the middleware is to improve the interaction between mobile clients and web services, with the support of the cloud platforms that used to improve the scalability and reliability of the middleware.

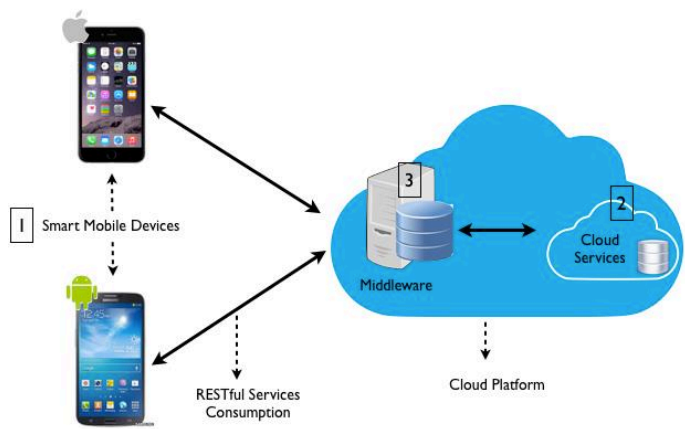

Figure 2: Middleware Architecture

#### 1) Middleware Architecture

The Middleware architecture, as shown in Figure 2, consists of the following three components:

1. A Mobile device that has Internet access.

2. A web application that contains web services connected to a database to store the needed data.

3. A web application that represents the middleware contains web services consumption, data format and handlers libraries.

#### 2) Middleware Advantages

The Middleware advantages can be summarized as following:

- Small bandwidth usage for limited GPRS data communication because of the light communication with the mobile client
- Little chance for failure in unreliable wireless networks.
- Takes the load of retrieving the response from the Web service.
- Advanced security features can be explored as the middleware acts on behalf of the mobile client.
- This architecture brings more opportunities towards ensuring a more reliable communication with the Web service such as:
  o The middleware will most likely run on dedicated hardware that will justify the search of solutions to ensure some kind of state of the communication with a Web service.
  o Retry mechanisms can be explored in case of connection failures.
  o Some communication states can ensure the transparency to the mobile client, such as in the case of communication failures (mobile device to middleware or middleware to Web service), the middleware can retain the state of the overall communication and retry to continue when all parties come back online.

#### 3) Middleware Limitations

The Middleware limitations are:

- The increasing of the overall duration of a request to the Web service, but this can be enhanced by deploying the middleware and the actual Web service in the same network (Via LAN).
- The heavy HTTP communication from middleware to Web service will bypass the network firewall thus adding an extra boost in the communication.
- The system might consume some time to process the data format on the middleware, but it may be lost because two communications lines have to be prior established and maintained.

### B. Mobile Agents (MA)

Mobile Agents are used to represent the user in wireless networks. They perform the requests using the user arguments and return the results to the user if s/he is online or the next connecting time, the user receives the results.

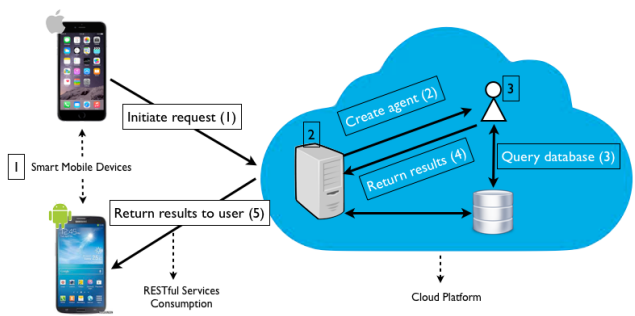

Figure 3: Mobile agent Architecture and the communication process between its components

### 1) MA Architecture

Mobile agents architecture as shown in Figure 3, consists of the following components:

1. A Mobile device that has Internet access.
2. A Web application that contains web services connected to a database to store the needed data.
3. A Mobile Agent that represents the user created by the system which consumes the services repository to get the response.

### 2) MA Advantages

Mobile agents advatnages can be summarized as following:

- Overcome the disconnection problem, it enriches the RESTFUL services using the mobile agent which represents the user in wireless networks.
- Doesn't need the user persistence in the entire session, so if the user disconnected due to network connection failure or empty battery, the user can get the results in the next time s/he connect to the system again.

### 3) MA Limitations

Mobile agents limitaions can be discussed as following:

- Agent gets and invokes only one service from the service repository.
- Multi-agents are needed in parallel to measure using parallelism on the response time if the user requested more than a service at the same time.

- The security layer is needed to be added to the request and the response to be used safely in confidential applications.

### VI. Results and Analysis

Different architectures are implemented based on the mentioned two approaches in Section V. Each architecture is concerned with a different case as summarized in Table 1.

| Reference | Approach | | Service Type | | Cache | |
|---|---|---|---|---|---|---|
| | Middle ware | Mobile Agent | Rest | SOAP | Mobile | Server |
| Reliable consumption of web services in a mobile-cloud ecosystem using REST [1] | Yes | No | Yes | No | Yes | Yes |
| Consuming Web Services on Mobile Platforms [5] | Yes | No | No | Yes | No | No |
| Enhance the Interaction Between Mobile Users and Web Services using Cloud Computing [9] | Yes | No | Yes | Yes | Yes | Yes |
| Wireless Semantic RESTFUL Services Using Mobile Agents [19] | No | Yes | Yes | No | No | Yes |

Table 1: The Comparison summary of Reliable web services approaches

In [1], [5], and [9] the authors applied the middleware approach, but each of them tailored the system with different features according to incases. In [1], the authors where interested in the Synchronization concept between cloud data and mobile device data. They followed the CAP Theorem by choosing the Availability and Partition tolerance as main points with slight consistency (Session Consistency).

The caching module was created in the middleware and the mobile device, the middleware caching is responding with the last saved results in case of cloud service unavailability, and the cache in the mobile device is used for showing the last results in case of internet connection unavailability. A notification mode was saved in the middleware to notify the mobile device once it becomes online again. In [5], the author is concerned with overcoming the heavy SOAP requests by converting them to lightweight requests.

The request flows as following: the mobile constructs the intended request with binary protocol or REST then sends it to the middleware that reconverts it to its original SOAP request and send it to the intended service. Regarding the response flow, the service response result is sent to the middleware to convert it to the equivalent lightweight format like: JSON then respond with it to the mobile device. The failure is overcame by creating a module

*2015 IEEE Seventh International Conference on Intelligent Computing and Information Systems (ICICIS'15)*to retry the request invocation based on the saved state when both cloud service and mobile device are online.

In [9], the author focused on two main points. The first is converting the heavy request to lightweight one, such that in the case of SOAP service, the middleware converts the SOAP to REST request then converts the heavy XML response result to JSON format.

The second is creating a caching module in both middleware and mobile device. The cache in the middleware is responding with the last saved results in case of cloud service unavailability, and the cache in the mobile device in order to show the last results in case of the internet connection unavailability. Some requests were connected directly to the cloud service to optimize the request and the response time.

In the Mobile Agent Approach [19], the author used the REST requests and constructed the agent once the mobile device connects to the cloud service.

The agent executes the request then either sends the response result to the mobile device if it was available or caches the results till the mobile reconnects to the service next time.

## XI. Conclusion and Future work

Achieving the reliable web service consumption is tended to two approaches: Middleware approach and Mobile Agent (MA) approach. Both of approaches focus on ensuring the request execution under the communication limitations and services temporary unavailability, such that each of them has tailored features to achieve the reliability.

The Middleware approach is more concerned with converting the heavy request to lightweight one before sending it in order to minimize the request time and communication overhead. Furthermore it depends on the retransmit mechanism when occurs a disconnection between the mobile device and cloud services. The middleware can be stay alive even if the cloud services aren't available, because of the separation between the middleware and the cloud services. Accordingly extra resources are needed to build the middleware architecture components, and more time is consumed for these additional operations.

The MA approach is interested in the request execution continuation even if the disconnection between the mobile device and the cloud service is occurred. The MA saves the result to respond with it in the next request from the same client.

These approaches used the caching mechanism to be able to resend the failed requests and retrieve the succeeded requests result when the mobile device reconnects.

Future work will be invested in implementing an hybrid approach that uses the advantages of middleware and mobile agent approaches. The proposed approach mainly concerns with the REST web service, ensures the low overhead communication, and keeps track for the mobile intermittent wireless connectivity state. Their tests will be conducted to measure the performance in terms of service response time, response size, network bandwidth, and connection error detection and correction. Results will be compared with the results of middleware and mobile agent approaches separately.

## XII. References

[1] R. K. Lomotey and R. Deters, "Reliable consumption of web services in a mobile-cloud ecosystem using REST," Proc. - 2013 IEEE 7th Int. Symp. Serv. Syst. Eng. SOSE 2013, pp. 13–24, 2013.

[2] J. H. Christensen, "Using RESTful web-services and cloud computing to create next generation mobile applications," Proc. 24th ACM SIGPLAN Conf. companion Object oriented Program. Syst. Lang. Appl., pp. 627–634, 2009.

[3] Y. He, O. S. Salih, C.-X. Wang, and D. Yuan, "Deterministic process-based generative models for characterizing packet-level bursty error sequences," Wirel. Commun. Mob. Comput., no. February 2013, pp. 421–430, 2009.

[4] W3C, "Web Services Architecture", 2004. Last accessed on October 2015, Available at: http://www.w3.org/TR/ws-arch/.

[5] A. COBARZAN, "Consuming Web Services on Mobile Platforms," Inform. Econ., vol. 14, no. 3, pp. 98–105, 2010.

[6] M. Chen, D. Zhang and L. Zhou, 'Providing web services to mobile users: the architecture design of an m-service portal', *International Journal of Mobile Communications*, vol. 3, no. 1, p. 1, 2005.

[7] S. McFaddin, D. Coffman, J. H. Han, H. K. Jang, J. H. Kim, J. K. Lee, M. C. Lee, Y. S. Moon, C. Narayanaswami, Y. S. Paik, J. W. Park, and D. Soroker, "Modeling and managing mobile commerce spaces using RESTful data services," Proc. - IEEE Int. Conf. Mob. Data Manag., pp. 81–89, 2008.

[8] J. Kleimola, 'A RESTful Interface to a Mobile Phone', *ResearchGate*, 2008. [Online].Available:http://www.researchgate.net/publication/228974867_A_RESTful_Interface_to_a_Mobile_Phone. [Accessed: 18- Nov- 2015].

[9] A. M. Gonsai and R. R. Raval, "ORIENTAL JOURNAL OF Enhance the Interaction Between Mobile Users and Web Services using Cloud Computing," pp. 1–9, 2014.

[10] V. Stirbu, "A RESTful architecture for adaptive and multi-device application sharing," Proc. First Int. Work. RESTful Des. - WS-REST '10, p. 62, 2010.

[11] H. H. H. Han, S. K. S. Kim, H. J. H. Jung, H. Y. Yeom, C. Y. C. Yoon, J. P. J. Park, and Y. L. Y. Lee, "A RESTful Approach to the Management of Cloud Infrastructure," 2009 IEEE Int. Conf. Cloud Comput., pp. 139–142, 2009.

[12] B. Sletten, "The REST architectural style in the Semantic Web," JavaWorld.com, April 2009. Last accessed on October 2015, Available: http://www.javaworld.com/javaworld/jw-04-2009/jw-04-rest-series-4.html.

[13] R. Lomotey and R. Deters, 'Using a Cloud-Centric Middleware to Enable Mobile Hosting of Web Services', *Procedia Computer Science*, vol. 10, pp. 634-641, 2012.

[14] E. a Brewer and U. C. Berkeley, "Inktomi at a Glance Distributed Systems Understanding Boundaries Where ' s the state ? ( not all locations are equal ) Santa Clara Cluster Delivering High Availability," Networks, vol. 19, pp. 1–12, 2000.

[15] S. Gilbert and N. Lynch, "Brewer's conjecture and the feasibility of consistent, available, partition-tolerant web services," ACM SIGACT News, vol. 33, no. 2, p. 51, 2002.

[16] S. Simon, "Brewer ' s CAP Theorem," pp. 1–6, 2000.

[17] D. Pritchett, "Base: an Acid Alternative," Queue, vol. 6, no. 3, pp. 48–55, 2008.227